# Architecting Strategic Influence: Operationalising the UXR Point of View Framework for Research Function Maturity

Rohini Singh
Admiral Group
rohini.singh50@admiralgroup.co.uk

Renée Barsoum
Admiral Group
renee.barsoum@admiralgroup.co.uk

**ABSTRACT**

This case study illustrates that the systematic application of the User Experience Research (UXR) Point of View (POV) framework serves as an effective operational scaffolding for a UXR function undergoing the critical transition from incubation to maturity. By assimilating structured 'Offensive' and 'Defensive' strategies, the presented Playbook equips UXR leaders with an adaptable toolkit to systematically navigate common institutional barriers, such as stakeholder bias, reactive tasking, and insight fragmentation. By pre-emptive and purposeful application of growth strategies, the likelihood of the research function establishing itself as a strategic partner capable of delivering evidence-based, actionable perspectives is significantly enhanced. The analysis demonstrates how this deliberate, Playbook-driven maturity strategy empowers research functions to move beyond tactical execution and directly shape long-term business strategy.

## 1. INTRODUCTION

User Experience Research (UXR) is increasingly recognized as a strategic discipline in its own right. Positioned at the intersection of user needs, business priorities, and organisational cognition, UXR functions are uniquely equipped to shape product direction, articulate uncertainty, and provide evidence-based directionality for long-term decision-making. However, establishing a new UXR function within a large, mature organization presents significant systemic friction. Nascent teams frequently confront structural and cultural impediments, including ad-hoc data practices, operational bottlenecks, stakeholder bias, and the persistent risk of being relegated to a reactive, service-oriented role [1].

This case study argues that cultivating an influential Research Function demands a sophisticated operational strategy that goes far beyond the execution of rigorous user studies. The UXR Point of View (POV) framework provides such scaffolding by supplying a structured approach for transforming raw data into shared organisational understanding. Through intentional deployment of the framework's Offensive and Defensive strategies, UXR leaders can architect a purposeful maturity pathway that accelerates the ascent of the research function to a strategic partner.

This maturity path parallels evolution in adjacent disciplines. Just as data organisations have progressed from descriptive reporting toward prescriptive influence [2], primarily by overcoming cultural rather than technical barriers. UXR functions must similarly cultivate the conditions for strategic authority. The POV pyramid provides a repeatable, diagnostic structure that enables research leaders to anticipate growth-stage friction and to tailor both Offensive plays (e.g., embedding insights in decision rituals, cultivating visual strategic frameworks) and Defensive plays (e.g., mitigating stakeholder bias, defending sampling strategies) [3]. When deployed iteratively, these Plays enable a function to maintain strategic continuity regardless of scale or organisational turbulence.

## 2. WORK IN THE INDUSTRY

Organizational psychology and design management literature suggest a persistent challenge in scaling knowledge work functions such as UXR. Nascent teams frequently struggle with the dual mandate of knowledge work: the need to balance rigorous methodological processes with the continuous delivery of strategic outcomes [4]. Early organizational pressures often force knowledge work teams into a path of either a strict process focus or outcome focus, with neither approach yielding the institutional trust required for strategic maturation [4]. Attempts to formally measure research productivity further show that relying on purely quantitative metrics often inadvertently incentivize tactical volume over strategic influence [5].

Several industry efforts have attempted to make the scaling of a UXR function more systematic and actionable for UXR leaders. One prominent approach is the use of standardized diagnostic models, such as Nielsen Norman Group's UX-Maturity Model, or Jared Spool's 5 Stages, which categorize a function's maturity and act as barometers for organizational integration [6]. These models assess the UXR functions across multiple axes, including leadership buy-in, operational standards, resource allocation, capacity building, and cross-functional integration [7]. Maturity stages generally evolve through recognizable phases: Emerging, Developing, Established, and Pioneering [8]. While these provide useful concepts for assessment, practitioners increasingly recognize the need to shift toward operational frameworks.

Efforts like the ResearchOps (ReOps) framework argue that scale is achieved by systematizing infrastructure, which alleviates logistical friction and improves researchers' strategic influence [7]. Another relevant effort involves synchronizing research rhythms with the broader business; frameworks like Scaled Agile (SAFe) embed Continuous Exploration into planning cycles to transform UXR into a predictive function [1]. The Maze 2025 Research Maturity report introduces a new dimension and advocates for scaling via democratization and AI, empowering non-researchers to handle tactical inquiries so the core team can focus on high-leverage questions [10].

Despite these advancements, there remains a gap in providing UXR leaders with practical, structured methodologies tailored to the daily cognitive and strategic transitions required for scaling. Strategic frameworks, such as those proposed by Emma Boulton, highlight four critical themes required for this maturity: Efficiency (standardizing infrastructure to enable scale), Visibility (ensuring research roadmaps are transparent and actionable), Quality (maintaining rigor as research is democratized), and Connectedness (breaking down silos to build unified organizational knowledge) [11]. These existing approaches offer broad principles without actionable steps for implementation.

Our proposed approach addresses this gap by introducing a set of structured Offensive and Defensive strategies that help research functions systematically navigate the cognitive hierarchy of the enterprise. This framework complements existing research by offering a hands-on methodology that enables UXR teams to mitigate reactive tasking and evolve from data reporting into Situational Awareness by utilizing the UXR Point of View [12].

## 3. METHODOLOGY

To systematically scale the nascent UXR function, a structured, three-step iterative methodology is deployed. This iterative process is designed to dynamically manage organizational complexity, resolve friction, and ensure the function's structural continuity toward becoming a strategic partner to the business. The development of growth strategies align closely with the structural mechanics of the UXR POV framework.

- **Challenge Identification:** We began with rigorous operational audits to identify structural blockers, whether cultural, political, or infrastructural, preventing the research function from advancing to the next stage of maturity. These were assimilated through multiple workshops involving research practitioners as well as stakeholders. The blockers identified during the challenge audits map directly against the hierarchical layers of the UXR POV Pyramid.
- **Strategy Development and Deployment:** Next, custom growth strategies were formulated. These correspond directly to the POV framework's concept of 'Plays' which are actionable sets of tactics designed to either clear organizational paths (Offensive Plays) or protect research rigor and boundaries (Defensive Plays). By assigning specific ownership and targeted Objectives and Key Results (OKRs) to autonomous sub-teams, the leadership generates institutional buy-in and maintains operational agility.
- **Measurement and Progress Checks:** The final phase establishes robust mechanisms to evaluate systemic impact. Metrics designed to track both the internal operational health of the function and its footprint on business decision-making. Viewed through the POV Framework, they track the realization of the overarching strategic narrative.

We utilized data from progress checks and metrics tracking to iteratively apply this process. A continuous feedback loop enabled research leadership to dynamically adjust strategic maneuvers and ensure the function maintains structural continuity toward its ultimate strategic mandate.

The iterative application of the POV framework at distinct growth stages acts as a catalytic scaffolding mechanism. It lends a powerful analytical and diagnostic lens, enabling UXR leaders to preemptively navigate organizational complexity.

## 4. THE STRATEGIC UXR FUNCTION PLAYBOOK

While the preceding methodology outlined the process of identifying organizational friction and building strategies, the Playbook operationalizes the strategies through a system of targeted 'Plays' and 'Cards'.

Anchoring the Playbook is a unifying principle, articulated through an overarching POV Narrative: *"A research function becomes strategic when it intentionally builds the structures and alignment needed to mature its practice and consistently translate user evidence into organisational impact."* This narrative serves as the North Star for the research function. It reinforces the premise that organisational maturity cannot be achieved merely by increasing the volume or methodological rigor of studies. Instead, maturity requires the deliberate, architectural construction of operational alignment across the organisation.

Every Play and Card within the UXR Playbook is engineered to advance the research function along this narrative arc. UXR leaders can utilize the Playbook to deploy tactics tailored to the unique environmental constraints of the research function's maturity level.

### 4.1 Defining the Scope of Included Plays

Across the maturity stages, a wide range of strategies were developed and applied to create purposeful growth and development of the UXR function. All of these contributed, in different ways, to strengthening the team's methodological capability and ability to articulate a strategic POV. However, several of these strategies, such as onboarding processes, administrative streamlining, operations and tooling, can be readily derived from established organisational best practices.

To keep the focus of this paper on the challenges unique to the growth of a UXR function, Plays that address UXR distinctive complexities, such as business alignment, methodological rigor, and the development of a coherent, organisationally resonant POV, are included in this case study. These Plays reflect the structured POV strategies and offer meaningful value to readers seeking to understand the evolution of UXR function.

### 4.2 Maturity Level 0: Incubation

**Goal:** Define the structural mandate to ensure a strong Foundation of the POV Pyramid.

The Incubation stage represents the most strategically fragile point in the evolution of a research function. At this early juncture, the team operates with constrained headcount, minimal institutional awareness, and uneven stakeholder expectations. Additionally, researchers encounter foundational frictions such as non-existent data infrastructure, ad-hoc tasking, and a lack of cross-functional trust.

The operational model at this stage functions analogously to a point guard on a basketball team. The UXR leader acts as the primary playmaker, maintaining a high-level awareness of the playing field, managing inbound executive requests, and making rapid yet deliberate decisions about where to distribute the ball (research effort) to maximise visibility, credibility, and impact. Like a point guard tasked with establishing tempo, spacing, and early scoring opportunities, the UXR lead is responsible for controlling the research rhythm, organisational attention, and ensuring that every possession (project) advances the broader strategic narrative of the function.

#### Play 1.1: Business Alignment and Foundations

Establishing the principles of engagement is critical to prevent a nascent research team from being defined as a reactive service desk. This play introduces UXR leaders to the foundational alignment mechanisms required for establishing a credible UXR function, address any interpretive gaps in the organization, and dictate operational cadence. The cards in this play (Table 1.1) highlight the early activities required to standardize terminology, formalize their strategic mandate, align stakeholder expectations, and codify internal workflows, thereby building a defensible operational foundation.

| Card Title | Challenge | Type | Quote | Guidance |
| --- | --- | --- | --- | --- |
| **Establishing First Definitions & Research Vocabulary** | Lack of a shared language leads to semantic drift and misaligned expectations. | Be Ready (Defensive) | *"Shared language is the first infrastructure of impact."* | Construct a shared language across stakeholders. Co-define key terms (e.g., user vs market research; behavioural vs attitudinal data, etc.) through facilitated sessions and formalise these in artefacts. By reinforcing a consistent vocabulary in plans, readouts, and strategy documents, researchers lay the linguistic groundwork required for an impactful POV. |
| **Mission & Mandate Definition** | Research being treated as a service rather than a strategic partner. | Be Ready (Defensive) | *"The essence of strategy is defining what not to do."* | Co-create a clearly articulated mission with both researchers and stakeholders for institutional buy-in. The mandate anchors UXR function in business objectives and reframes the team as strategic partner. |
| **Stakeholder Alignment & Expectation Setting** | Misalignment between stakeholder expectations and UXR's POV driven approach. | Best Practice (Offensive) | *"A Point of View is a shared journey toward evidence-based decision making."* | Alignment requires shifting stakeholder mindsets away from data delivery toward collaborative evolution of a POV. Introducing the POV Pyramid early helps stakeholders understand how raw data becomes wisdom, reducing friction at delivery. |

| Education & Promotion | Lack of organizational awareness of UXR's purpose. | Best Practice (Offensive) | *"Repetition builds recognition."* | Sustained, organisation-wide educational outreach through roadshows, introductory sessions, and presentations. Repetition of message and visibility gradually reshape organisational cognition, enabling stakeholders to internalise UXR's strategic remit. |
|---|---|---|---|---|

Table 1.1 Cards corresponding to Play 1.1

**Play 1.2: Establishing Methodological Rigor and Quality Standards**

Institutional trust in a newly formed research function is highly fragile and relies heavily on the capacity of researchers to act as recognized, authoritative knowers within a socio-technical system. Ensuring high-quality research outputs in the early stages of a function's development is crucial for establishing credibility. By enforcing quality controls and standardized rituals, this Play (Table 1.2) enables UXR leaders to ensure that the function produces strong outcomes.

| Card Title | Challenge | Type | Quote | Guidance |
|---|---|---|---|---|
| **Insight Quality & Peer Review Standards** | Differing standards of research rigour. | Best Practice (Offensive) | *"Quality is not an act, it is a habit."* | Set up quality expectations by implementing formal peer-review protocols across the research lifecycle, from study design to synthesis. Codify a unified definition of an actionable insight. |
| **Establishing Research Rituals** | Inconsistent or fragmented processes weaken reliability of outputs. | Best Practice (Offensive) | *"Rituals turn scattered projects into a unified strategic engine."* | Create repeatable rituals (e.g., scoping workshops, bi-weekly syncs, reviews, playbacks) that render the research process visible, standardised, and aligned with the pyramid's layers. These rituals function as strategic touchpoints, enabling consistent expectations and enabling cumulative organisational learning. |
| **Mixed-Methods Activation** | Relying exclusively on single-method evidence limits strategic impact. | Best Practice (Offensive) | *"Triangulation is transformation."* | Triangulation ensures the UXR function generates a highly resilient and persuasive POV that commands executive confidence. UXR leaders activate mixed-method approaches through rituals and reviews to ensure researchers capitalize different forms of evidence. |
| **Addressing Inherited Practices** | Disparate research habits risk organizational trust. | Be Ready (Defensive) | *"Technical debt in human form"* | Audit the diverse practices imported by individual researchers, deprecating incompatible practices and integrating best-in-class techniques into the team's operations, transforming a fragmented group into a unified function. |

Table 1.2 Cards corresponding to Play 1.2

**Play 1.3: Establishing an Impact-Driven Culture**

A strategic research function requires more than methodological rigor: it requires a culture that consistently orients itself toward creating impact. This Play (Table 1.3) introduces UXR leaders to cards that engineer an internal culture that systematically expands the research function's strategic footprint across the enterprise.

| Card Title | Challenge | Type | Quote | Guidance |
|---|---|---|---|---|
| **Bringing the Right Talent** | The skill gap impacts the business's trust in the UXR function. | Best Practice (Offensive) | *"Sequence your city: builders first, shapers next, dwellers when foundations are strong."* | Define hiring strategy to ensure the cultural centre of gravity is set by strategically minded researchers. 'City Builders' (senior generalists) create the foundational culture; 'City Shapers' then add methodological diversity; 'City Dwellers' (junior researchers) enter only once the practice is stable enough to support growth. |
| **Cross-Functional (XFN) Collaboration** | Research impact is weakened without cross functional relationships. | Best Practice (Offensive) | *"Great solos don't win symphonies"* | Collaboration is a core cultural competency for an impact driven UXR function. Through proactive embedding in decision cycles, researchers ensure that insights influence roadmaps. |

| Coaching from Data to Wisdom | If researchers don't own the POV, strategic impact diminishes. | Best Practice (Offensive) | *"Practice makes perfect"* | Introduce structured coaching designed to elevate the analytical maturity of the research team. By clarifying the UXR function's strategic mandate, leaders ensure that every researcher establishes and defends a compelling, strategically relevant POV. |
|---|---|---|---|---|

Table 1.3 Cards corresponding to Play 1.3

**Play 1.4: Architecting the Scaling Roadmap**

As the nascent research function solidifies its organisational mandate, secures its methodological rigor, and cultivates an impact driven culture, UXR leaders must also simultaneously focus on establishing a purposeful blueprint that will shape the team's growth trajectory. This Play (Table 1.4) operates as the deliberate mechanism to propel the overarching POV narrative forward, actively designing the function's future state and securing the executive sponsorship required to cross the threshold into Scaling.

| Card Title | Challenge | Type | Quote | Guidance |
|---|---|---|---|---|
| **Accelerate Growth Through Pet Projects** | Leveraging researcher passion to build long term strategic capability | Best Practice (Offensive) | *"Incubate your POV through strategic pet projects."* | Establish a formalised mechanism for researcher-led exploration through 'Pet projects'. Aligning researcher's personal curiosity with organisational strategy creates a proactive learning culture, mitigates researcher burnout, and transforms individual exploration into organisational advantage. |
| **Capability Mapping and Gap Analysis** | Scaling efforts risk compounding hidden infrastructural bottlenecks. | Best Practice (Offensive) | *"What got you here won't get you there."* | Systematically assess strengths and deficits of the team's current operational state (e.g., specialised talent, tooling, methodological breadth), against the capabilities required at the next maturity tier. Create structured capability maps that define the roles, systems, and practices that must be introduced or strengthened to support sustainable scaling. |
| **Strategic Evangelism and Impact Broadcasting** | Without broadcasting successes, strategic value remains invisible, stalling cross-functional momentum and executive buy-in. | Best Practice (Offensive) | *"Under communication by a factor of ten, or even a hundred, is one of the most common errors in change management."* | Engineer a deliberate internal communication and marketing plan to broadcast the UXR function's emerging impact. Tailor strategic narratives to different stakeholder groups, clarifying how UXR advances their objectives. By consistently celebrating impact across the business, UXR leader translates isolated tactical wins into a coherent story of creating compelling POVs, generating the demand and political capital required to support the Scaling roadmap. |

Table 1.4 Cards corresponding to Play 1.4

**4.3 Maturity Level 1: Scaling (Infrastructure & Efficiency)**

**Goal:** Navigating volume and elevating business awareness

The successful execution of Level 0 operational plays inevitably triggers a surge in research requests. The transition to Scaling stage begins as a deliberate, architected progression in which UXR leaders proactively re-apply the challenge-identification, strategy-development, and measurement cycle that underpins the POV framework to anticipate and neutralise emerging friction ahead of time.

The primary objective during the Scaling stage evolves from proving foundational value to rapid outcome delivery while protecting established quality standards. The centralized triage model (point guard) is adapted accordingly to avoid operational bottlenecks. Conversely, the common reflex to fully decentralise the team by embedding researchers into product squads creates a new vulnerability of insight fragmentation, failing to feed back into the overarching POV Narrative.

We evolve our operating model into a coordinated, domain-aligned structure, much like Zone defence in basketball. The UXR team remains centrally governed as one cohesive analytical unit, preserving the quality standards, rituals, and momentum established in Level 0. Research leads are assigned to specific strategic domains (their zones) who, like zone defenders, maintain awareness of their domain, anticipate incoming needs, and call out emerging risks or opportunities. This allows capacity to be surged or redistributed across zones as new executive requests come in or priorities shift. The UXR function builds deeper local expertise (strategic zones) while preserving agility.

**Play 2.1: The Lean Research Play**

This Play (Table 2.1) is designed to increase efficiency while preserving research integrity through standardised rituals, proportional methods, and embedded quality governance. It operationalises a set of proportionate, repeatable approaches that shorten research cycle times while preserving the validity of insights.

| Card Title | Challenge | Type | Quote | Guidance |
|---|---|---|---|---|
| **Lean Research Methods** | As research demand outpaces capacity; teams oscillate between over-engineered studies and superficial fast work. | Best Practice (Offensive) | *“Not every question requires a cathedral of methods.”* | Formalise tiered research patterns (micro-tests, lean sprints). Embed rapid-synthesis rituals (same-day debriefs, rolling affinity, evidence snapshots) aligned to the POV Pyramid layers. Ensure right sizing is tied to decision risk, triangulating when the decision risk justifies it. |
| **Insight Quality & Peer Review Standards** | Increasing velocity creates inconsistent standards of rigour and weakens institutional trust. | Best Practice (Offensive) | *“Quality is not an act; it is a habit.”* | Utilise the established peer-review practices to coach researchers towards lean research. Reviews serve as a quality stabiliser in a high throughput environment. |
| **Tooling & Templates** | Repetitive administrative tasks divert researchers’ bandwidth. | Best Practice (Offensive) | *“Standardization is the necessary foundation for tomorrow's improvement."* | Create a centralized library of templates (e.g. scoping, research plan, discussion guide, reports) and automated recruitment pipelines, stripping away the logistical friction. |
| **AI Augmented Synthesis** | As qualitative data volumes scale, manual transcription and thematic tagging become bottlenecks. | Best Practice (Offensive) | *“Automation applied to an efficient operation will magnify the efficiency.”* | Operationalise the strategic integration of AI-driven analysis tools into the synthesis workflow. Enforce rigorous audits on AI-generated outputs, to ensure that the integrity and validity of the POV is not compromised. |

Table 2.1 Cards corresponding to Play 2.1

**Play 2.2: Systematizing and Quantifying Impact**

This Play introduces UXR leaders to proactively establish the impact infrastructure required for a scaling UXR function to thrive. The acquisition of new headcount, advanced tooling, and expanded operational scope (secured via Play 1.4) transforms the UXR function into a significant capital investment. This Play equips UXR leaders with the mechanisms to formalise how business value is defined, measured, and communicated. By establishing consistent impact attribution across risk reduction, revenue influence, operational efficiency, and customer-centric outcomes, leaders ensure that UXR research efforts remain strategically synchronised and continuously reinforce the functions’ overarching POV Narrative. This Play systematizes the definition, integration, and longitudinal tracking of business value delivered by the UXR function.

| Card Title | Challenge | Type | Quote | Guidance |
|---|---|---|---|---|
| **Formulating the Impact Framework** | Without a shared, standard way to measure impact, leaders cannot see or prove the research function’s ROI. | Best Practice (Offensive) | *“Measure what matters.”* | Design a standardized Impact Measurement Framework that transforms the UXR function’s outcomes into quantifiable business value. This enables UXR leaders to ensure that every study evaluates success against enterprise priorities, enabling the team to build a coherent, defensible narrative of cumulative strategic impact. |
| **Integrating Impact into the Research Lifecycle** | Retrospective impact assessment reduces meaningful influence | Best Practice (Offensive) | *“To begin with the end in mind means to start with a clear understanding of your destination.”* | Embed the Impact Framework directly into core workflows by requiring research plans, scoping templates, and intake briefs to articulate its intended business impact upfront. This ensures studies begin with a |
| | | | | clear value target and that researcher capacity is spent only on work engineered to drive measurable strategic value. |
| **Operationalizing the Impact Tracker** | Without a centralized recording, the function's impact risks fading from executive memory. | Be Ready (Defensive) | *"Without data, you're just another person with an opinion."* | Create a continuous ROI reporting through a centralized, longitudinal Impact Tracker. Logging how research drives shift in product and business decisions, constructs a data-driven narrative of business impact. |

Table 2.2 Cards corresponding to Play 2.2

## 5. MEASURES AND OUTCOMES

To assess the effectiveness of the strategies introduced and the UXR function's progression across the maturity stages, we assessed performance across 3 core dimensions: Researcher Experience, Organisational Impact, and Functional Health. Our measures were purposefully focussed on outcomes rather than outputs, ensuring that progress was evaluated not by the volume of studies produced but by the degree to which the function strengthened its influence, supported strategic decision-making, and aligned with business priorities. Establishing these measures early in the process helped reinforce the behaviours needed for long-term success and provided leadership with a clear view of the function's structural health as it navigates toward maturity.

| Performance Domain | Measure | Rationale | Results |
|---|---|---|---|
| **Researcher Experience** | Practitioner engagement, role satisfaction, and cognitive autonomy. Tracked longitudinally via quarterly engagement surveys and retention data. | Foundational measures of a resilient function. These measures enable the UXR leaders to track and ensure the operational architecture empowers rather than exhausts the team. | Interventions successfully fostered high psychological safety and strategic ownership. Notably, the research function did not simply stabilise, it outperformed parallel teams across the same organisational measures. Further validated by exceptional retention during hyper-growth (scaling from 3 to 20 practitioners in 18 months with near-zero voluntary attrition). |
| **Organisational Impact** | Penetration of findings into the organizational strategic membrane tracked via successful roadmap insight integrations, strategic pivots, and quantifiable risk mitigation. | To validate the translation of research outcomes into business value. As the function scales, tracking these metrics enable UXR leaders to ensure research efforts actively aggregate to shape executive decision-making. | Insight absorption increased significantly from Incubation to Scaling stage. Defining 'zones' positioned more researchers to directly inform upstream product roadmaps. Multiple strategic studies yielded substantial developmental cost savings by pivoting initiatives before engineering capital was expended. |
| **Functional Health** | The business's structural and financial engagement with the team. Tracked via resource allocation, headcount expansion, and intake compliance. | Institutional trust is an empirical reality demonstrated through financial investment, validating perceived strategic influence among executive stakeholders. | Executive leadership transitioned from viewing UXR as discretionary overhead to a critical capital investment (funding a nearly 7x team expansion). Cross-functional partners shifted engagement from downstream validation to upstream strategy. |

## 6. DISCUSSION

The operational scaffolding detailed in this case study illustrates the UX Research function's intentional transition from foundational Incubation to Scaling. The strategic evolution from a highly centralized point guard model to a semi-zoning architecture, coupled with rigorous lean methodologies and formalized impact tracking, successfully supported the growth of the UXR function.

In this case study, we demonstrated that the journey from an isolated, ad-hoc research practice to an influential UXR function requires the intentional, systematic application of a growth strategy. By utilizing the UXR POV framework, UXR leaders can diagnose their current maturity, anticipate structural bottlenecks, and deploy targeted Plays to systematically construct their organizational influence. The outcomes observed through the application of this framework to the initial stages of maturity prove that when UXR is treated as an intentional structural discipline, it possesses the capacity to influence business as a strategic partner.

## 7. PLANNED FUTURE WORK: AI-ASSISTED STRATEGY ARCHITECTURE

As the UXR function advances through the Scaling stage, we are leveraging the POV Framework to anticipate the frictions and requirements inherent in reaching the 'Established' and subsequently 'Pioneering' stages of maturity. One of our main focus areas will be developing strategies to elevate the researchers' from having domain knowledge into specialized strategic advisors. We aim to explore how the UXR function can collaborate effectively with the Business Strategy group and directly inform formation of business strategy. Furthermore, the framework will be deployed to address the critical organizational friction of knowledge silos. We plan to engineer formal collaboration with adjacent insights functions, such as Data Science and Market Insights. The ultimate challenge to solve in this next phase is determining how to systematically merge these diverse, fragmented data sources to craft compelling POVs that eliminate organizational blind spots.

A critical secondary area of future work involves exploring the strategic application of Artificial Intelligence to architect the organizational playbook itself. While earlier scaling phases utilized AI as a tactical accelerant for operational efficiency, the upcoming challenge lies in elevating AI to act as a strategic co-pilot in organizational design. We will investigate how to leverage AI to dynamically generate the operational Plays and constituent Cards required to navigate our next growth step into the Established stage. By inputting the anticipated structural frictions, enterprise goals, and the constraints of the UXR POV framework, we aim to understand how AI can assist UXR leadership in authoring rigorous, context-specific strategies, pre-empt organizational bottlenecks, and draft the operational roadmap for maturity.